\documentclass{article}
\usepackage{spconf}

\usepackage{cite}
\usepackage[T1]{fontenc}
\usepackage{calrsfs}
\usepackage{amsmath, accents, amssymb, amsfonts, bm, mathtools, cases}
\usepackage{algorithm}
\usepackage{algpseudocode}
\usepackage[inline]{enumitem}
\usepackage{graphicx}
\usepackage{hyperref}

\usepackage{subcaption}

\usepackage{textcomp}
\usepackage{xcolor}
\usepackage{etoolbox}
\def\BibTeX{{\rm B\kern-.05em{\sc i\kern-.025em b}\kern-.08em
    T\kern-.1667em\lower.7ex\hbox{E}\kern-.125emX}}
\usepackage[capitalize]{cleveref}
\usepackage{url}

\usepackage{pgfplots}
\pgfplotsset{compat=newest}
\usepgflibrary[plotmarks]
\usepgfplotslibrary{fillbetween}
\usetikzlibrary{angles, arrows, arrows.meta, shapes, tikzmark, patterns}

\newtheorem{theorem}{Theorem}

\newcommand\given{{\mathbin{}\mid\mathbin{}}}
\newcommand{\IntegerP}{\mathbb{N}}
\newcommand{\IntegerPP}{\mathbb{N}_*}
\newcommand{\Real}{\mathbb{R}}
\newcommand\SetSymbol[1][]{
  \nonscript\,#1\vert \allowbreak \nonscript\,\mathopen{}}
\DeclarePairedDelimiterX\Set[1]{\lbrace}{\rbrace}%
{ \renewcommand\given{\SetSymbol[\delimsize]} #1 }
\DeclarePairedDelimiterX\norm[1]\lVert\rVert{\ifblank{#1}{\:\cdot\:}{#1}}
\DeclarePairedDelimiterX\innerp[2]{\langle}{\rangle}{#1
  \mathop{}\delimsize\vert\mathop{} #2}

\DeclareMathOperator{\Fix}{Fix}
\DeclareMathOperator{\linspan}{span}

\crefname{theorem}{Theorem}{Theorems}
\crefname{thmlisti}{Theorem}{Theorems}
\crefname{algo}{Algorithm}{Algorithms}
\crefname{figure}{Figure}{Figures}
\crefname{section}{Section}{Sections}

\newlist{thmlist}{enumerate}{1}
\setlist[thmlist]{label=\textbf{(\roman{*})}, ref=\thetheorem(\roman{*}), noitemsep}

\hypersetup{hidelinks}
\allowdisplaybreaks

\title{Dynamic Selection of p-Norm in Linear Adaptive Filtering via\\
  Online Kernel-Based Reinforcement Learning\vspace{-30pt}%
}

\name{}

\address{%
  \begin{minipage}{.7\textwidth}
    \begin{center}
      \textit{Minh Vu\qquad Yuki Akiyama\qquad Konstantinos Slavakis}\\[1ex]
      \small Tokyo Institute of Technology, Japan\\
      Department of Information and Communications Engineering\\
      Emails: \texttt{\{vu.d.aa, akiyama.y.am, slavakis.k.aa\}@m.titech.ac.jp}
    \end{center}
  \end{minipage}
}

\begin{document}
\ninept

\maketitle

\begin{abstract}
  This study addresses the problem of selecting dynamically, at each time instance, the
  ``optimal'' p-norm to combat outliers in linear adaptive filtering without any knowledge on
  the potentially time-varying probability density function of the outliers. To this end,
  an online and data-driven framework is designed via kernel-based reinforcement learning
  (KBRL). Novel Bellman mappings on reproducing kernel Hilbert spaces (RKHSs) are introduced
  that need no knowledge on transition probabilities of Markov decision processes, and are
  nonexpansive with respect to the underlying Hilbertian norm. An approximate policy-iteration
  framework is finally offered via the introduction of a finite-dimensional affine superset of
  the fixed-point set of the proposed Bellman mappings. The well-known ``curse of
  dimensionality'' in RKHSs is addressed by building a basis of vectors via an approximate
  linear dependency criterion. Numerical tests on synthetic data demonstrate that the proposed
  framework selects always the ``optimal'' p-norm for the outlier scenario at hand,
  outperforming at the same time several non-RL and KBRL schemes.
\end{abstract}


\section{Introduction}\label{sec:intro}

The least-squares (LS) error/loss (between an observed value and its predicted one) plays a
pivotal role in signal processing, e.g., adaptive filtering~\cite{sayed2011adaptive}, and
machine learning~\cite{Theodoridis.Book:ML}. For example, the least-mean squares (LMS) and
recursive (R)LS~\cite{sayed2011adaptive} are two celebrated algorithms in adaptive filtering
and stochastic approximation which are based on the LS-error criterion. Notwithstanding, LS
methods are notoriously sensitive to the presence of outliers within data~\cite{rousseeuw1987},
where outliers are defined as (sparsely) contaminating data that do not adhere to a nominal
data generation model, and are often modeled as random variables (RVs) with non-Gaussian heavy
tailed distributions, e.g., $\alpha$-stable ones~\cite{shao1993signal, miotto2016pylevy}. To
combat the negative effects of outliers, several non-LS criteria, such as least mean $p$-power
(LMP)~\cite{pei1994p-power, xiao1999adaptive, Kuruoglu:02, vazquez2012, chen2015smoothed,
  slavakis2021outlier} and maximum correntropy (MC)~\cite{Singh.MCC:09}, have been
studied. This work focuses on the LMP criterion, owing to the well-documented robustness of LMP
against outliers~\cite{Gentile:03}, while results on MC will be reported elsewhere.

This study builds around the classical data-generation model
$y_n = \bm{\theta}_*^{\intercal} \mathbf{x}_n + o_n$, where $n\in\IntegerP$ denotes discrete
time ($\IntegerP$ is the set of all non-negative integers), $\bm{\theta}_*$ is the $L\times 1$
vector whose entries are the system parameters that need to be identified, $o_n$ is the RV
which models outliers/noise, $(\mathbf{x}_n, y_n)$ stands for the input-output pair of
available data, where $\mathbf{x}_n$ is an $L\times 1$ vector and $y_n$ is real-valued, and
$\intercal$ denotes vector/matrix transposition. For an arbitrarily fixed $\bm{\theta}_0$, the
LMP algorithm~\cite{pei1994p-power} generates estimates $(\bm{\theta}_n)_{n\in\mathbb{N}}$ of
$\bm{\theta}_*$ according to the following recursion:
\begin{equation}
  \bm{\theta}_{n+1} \coloneqq \bm{\theta}_n + \rho p \lvert e_n\rvert^{p-2}e_n \mathbf{x}_n
  \,, \label{LMP}
\end{equation}
where $e_n \coloneqq y_n - \mathbf{x}_n^{\intercal} \bm{\theta}_n$, $\rho$ is the learning rate
(step size), and $p$ is a \textit{fixed}\/ user-defined real-valued number within the interval
$[1, 2]$ to ensure that the $p$-norm loss
$\lvert y_n - \mathbf{x}_n^{\intercal} \bm{\theta} \rvert^p$ is a convex function of
$\bm{\theta}$~\cite{pei1994p-power}. Notice that if $p = 1$ and $2$, then \eqref{LMP} boils
down to the classical sign-LMS and LMS, respectively~\cite{sayed2011adaptive}.

Intuition suggests that the choice of $p$ should be based on the distribution of $o_n$ in the
data-generation model. For example, if $o_n$ obeys a Gaussian distribution, then $p=2$ should
be chosen (recall the maximum-likelihood criterion). To enhance robustness against outliers,
combination of adaptive filters with different forgetting factors, but with the same fixed
$p$-norm, have been also introduced~\cite{vazquez2012}. However, it seems that an
\textit{online}\/ and \textit{data-driven}\/ solution to the problem of \textit{dynamically}\/
selecting $p$, \textit{without}\/ any prior knowledge on the distribution of $o_n$, which may
\textit{change with time,} is yet to be found.

This work offers a solution to the aforementioned open problem via reinforcement learning
(RL)~\cite{bertsekas2019reinforcement}; a machine-learning paradigm where an ``agent''
interacts with the surrounding environment to identify iteratively the policy which minimizes
the cost of its ``actions.'' More specifically, the well-known policy-iteration (PI)
framework~\cite{bertsekas2019reinforcement} of RL is adopted, because of its well-documented
merits (e.g., \cite{ormoneit2002kernel, Ormoneit:Autom:02, xu2007klspi}) over the alternative
RL frameworks of temporal-difference (TD) and Q-learning~\cite{bertsekas2019reinforcement},
especially for continuous and high-dimensional state spaces such as the one considered here. PI
comprises two stages at every iteration $n$: \textit{policy evaluation} and \textit{policy
  improvement}. At policy evaluation, the current policy is evaluated by a
$Q$-function~\cite{bertsekas2019reinforcement}, which represents, loosely speaking, the
long-term cost that the agent would suffer had the current policy been used to determine the
next state, whereas at the policy-improvement stage, the agent uses the $Q$-function values to
update the policy. The underlying state space is considered to be continuous and high
dimensional, due to the nature of the available data $(\mathbf{x}_n, y_n)$, while the action
space is considered to be discrete: an action is a value of $p$ taken from a finite grid of the
interval $[1,2]$.

\sloppy Deep neural networks offer approximating spaces for Q-functions, e.g., \cite{DDQN}, but
they may require processing of batch data (even re-training) during online-mode operation,
since they may face test data generated by probability density functions (PDFs) different
from those of the training ones (dynamic environments). Such batch processing inflicts large
computational times and complexity, discouraging the application of deep neural networks to
online learning where a small complexity footprint is desired.

To meet the desired computational complexity requirements, this study builds an approximate
(A)PI framework for \textit{online}\/ RL along the lines of kernel-based
(KB)RL~\cite{ormoneit2002kernel, Ormoneit:Autom:02, xu2007klspi, Bae:MLSP:11, Barreto:NIPS:11,
  Barreto:NIPS:12, Kveton_Theocharous_2013, OnlineBRloss:16, RegularizedPI:16,
  Kveton_Theocharous_2021, Wang_Principe:SPM:21}. Central to the API design is the construction
of novel Bellman mappings~\cite{bellman2003dp, bertsekas2019reinforcement}. The proposed
Bellman mappings are defined on a reproducing kernel Hilbert space (RKHS)
$\mathcal{H}$~\cite{aronszajn1950, scholkopf2002learning}, which serves as the approximating
space for the $Q$-functions. In contrast to the prevailing route in
KBRL~\cite{ormoneit2002kernel, Ormoneit:Autom:02, xu2007klspi, Bae:MLSP:11, Barreto:NIPS:11,
  Barreto:NIPS:12, Kveton_Theocharous_2013, OnlineBRloss:16, RegularizedPI:16,
  Kveton_Theocharous_2021, Wang_Principe:SPM:21}, which views Bellman mappings as contractions
in $\mathcal{L}_{\infty}$-norm Banach spaces (by definition, no inner product available), this
study introduces nonexpansive~\cite{HB.PLC.book} Bellman operators on $\mathcal{H}$ to
capitalize on the reproducing property of the inner product of
$\mathcal{H}$~\cite{aronszajn1950, scholkopf2002learning}, and to open the door to powerful
Hilbertian tools~\cite{HB.PLC.book}. This path offers also flexibility to the user to choose
any point from the potentially non-singleton fixed-point set of the nonexpansive Bellman
mapping, as opposed to the case of a contraction mapping which is known to have a unique fixed
point. A superset of the fixed-point set of one of the proposed Bellman mappings is designed
onto which the proposed API framework is based. To robustify the policy-improvement stage, the
well-known methodology of rollout~\cite{bertsekas2019reinforcement} is employed. Moreover, to
address the issue of the ``curse of dimensionality'' that arises naturally in online learning
in RKHSs, the proposed framework is equipped with the approximate linear dependency (ALD)
criterion~\cite{engel2004krls}. Note that~\cite{Bae:MLSP:11, Wang_Principe:SPM:21}, being along
the lines of TD and Q-learning, do not include any discussion on Bellman mappings.

Unlike the classical Bellman operators, where information on transition probabilities of a
Markov decision process is needed~\cite{bertsekas2019reinforcement}, the proposed Bellman
mappings need neither such information nor any training/offline data and past policies, but
sample and average the state space on-the-fly, at each $n$, to \textit{explore}\/ the
surrounding environment. This suits the current adaptive-filtering setting, where the presence
of outliers, with a possibly time-varying PDF, may render the information obtained offline or
from past policies outdated. As such, the proposed Bellman mappings fall closer
to~\cite{Ormoneit:Autom:02} than to studies which use training data collected beforehand
(offline), e.g., \cite{lagoudakis2003lspi, xu2007klspi, panaganti2022robust}.

It is worth stressing here that the proposed framework, together with its complementary
study~\cite{Yuki:ICASSP23}, appear to be the first attempts to apply RL arguments to robust
adaptive filtering. In contrast to \cite{Yuki:ICASSP23}, where the state space is the
low-dimensional $\Real^4$, this study considers the high-dimensional $\Real^{2L+1}$ ($\Real$ is
the set of all real numbers). This work constructs a finite-dimensional affine set as a
superset of the fixed-point set of one of the proposed Bellman mappings, as opposed to
\cite{Yuki:ICASSP23} where a potentially infinite-dimensional hyperplane is designed. To reduce
the size of the computational footprint of the proposed framework, ALD is used instead of the
random Fourier features (RFF)~\cite{RFF} in~\cite{Yuki:ICASSP23}. Finally,
rollout~\cite{bertsekas2019reinforcement} is employed for robustification, whereas experience
replay~\cite{ExperienceReplay} is applied in~\cite{Yuki:ICASSP23}.

Numerical tests on synthetic data demonstrate that the advocated framework \textit{always}\/
spots the value of $p$ that leads to ``optimal'' performance, \textit{without}\/ any knowledge
on the PDF of the outliers, which is intentionally made to be time varying. Due to space
limitations, any proofs, results on convergence analysis, and further numerical tests will be
reported in the journal version of this paper.

\section{Nonexpansive Bellman Mappings on RKHS\lowercase{s}}\label{sec:nonexp.Bellman}

\subsection{State-Action Space}\label{sec:state.action.space}

Following \eqref{LMP}, the state space is defined as the following continuous and
high-dimensional
$\mathfrak{S} \coloneqq \{ \mathbf{s} \coloneqq ( \mathbf{x}, y, \bm{\theta} ) \given
\mathbf{x}\in \Real^L, y\in \Real, \bm{\theta}\in \Real^L\} = \Real^{2L+1}$, where $\Real$
stands for the set of all real numbers. The action space $\mathfrak{A}$ is defined as any
finite grid of the interval $[1,2]$, so that an action $a\in \mathfrak{A}$ becomes any value of
$p$ taken from that finite grid. The state-action space is defined as
$\mathfrak{Z} \coloneqq \mathfrak{S}\times \mathfrak{A}$, and its element is denoted as
$\mathbf{z} = (\mathbf{s}, a)$.

Along the lines of the general notation in~\cite{bertsekas2019reinforcement}, consider now the
set of all mappings
$\mathcal{M} \coloneqq \Set{ \mu(\cdot) \given \mu(\cdot): \mathfrak{S} \to \mathfrak{A}:
  \mathbf{s} \mapsto \mu(\mathbf{s})}$. In other words, given a $\mu\in \mathcal{M}$,
$\mu(\mathbf{s})$ denotes the action that the ``system'' may take at state $\mathbf{s}$ to
``move to'' the state $\mathbf{s}^{\prime}\in \mathfrak{S}$. The set $\Pi$ of policies is
defined as
$\Pi \coloneqq \mathcal{M}^{\IntegerP} \coloneqq \Set{ (\mu_0, \mu_1, \ldots, \mu_n, \ldots)
  \given \mu_n\in \mathcal{M}, n\in \IntegerP}$. A policy will be denoted by $\pi\in
\Pi$. Given $\mu\in \mathcal{M}$, the stationary policy $\pi_{\mu} \in \Pi$ is defined as
$\pi_{\mu} \coloneqq ( \mu, \mu, \ldots, \mu, \ldots)$. It is customary for $\mu$ to denote
also the stationary policy $\pi_{\mu}$.

The one-step loss $g: \mathfrak{Z}\to \Real: (\mathbf{s}, a) \mapsto g(\mathbf{s}, a)$, which
quantifies the cost of transition from the current state $\mathbf{s}$ to the next one
$\mathbf{s}^{\prime}$ under action $a$, is defined for the current setting as:
\begin{align}
  g(\mathbf{s}, a) \coloneqq \vert y - \bm{\theta}^{\intercal} \mathbf{x} \vert ^a + \vert y
                 - \bm{\theta}^{\prime}{}^{\intercal} \mathbf{x} \vert \,, \label{one.step.loss}
\end{align}
where
$\bm{\theta}^{\prime} \coloneqq \bm{\theta} + \rho a \lvert y - \bm{\theta}^{\intercal}
\mathbf{x} \rvert^{a-2}(y - \bm{\theta}^{\intercal} \mathbf{x}) \mathbf{x}$, according to
\eqref{LMP}. Recall that action $a$ is defined to take certain values of $p$ from a
user-defined grid in $[1,2]$. The loss in \eqref{one.step.loss} is motivated by classical
adaptive filtering~\cite{sayed2011adaptive}; namely, the first term in \eqref{one.step.loss}
resembles the \textit{prior loss,} while the second one mimics the \textit{posterior loss.}
Only the prior loss is chosen to be affected by action $a$ in \eqref{one.step.loss}, because
$a$ applies to the current state $\mathbf{s}$ and not $\mathbf{s}^{\prime}$. The long-term loss
$Q: \mathfrak{Z}\to \Real: (\mathbf{s}, a) \mapsto Q(\mathbf{s}, a)$ quantifies the long-term
cost that the agent would suffer had the action $a$ been used to determine the next state of
$\mathbf{s}$.

\subsection{Novel Bellman Mappings}\label{sec:novel.Bellman.maps}

Central to dynamic programming and RL~\cite{bertsekas2019reinforcement} is the concept of
Bellman mappings which operate on $Q$-functions. Typical definitions are, e.g.,
\cite{Bellemare:16}, $\forall (\mathbf{s}, a)\in \mathfrak{Z}$,
\begin{subequations}\label{Bellman.maps.standard}
  \begin{align}
    (T_{\mu}^{\diamond} Q)(\mathbf{s}, a)
    & \coloneqq g( \mathbf{s}, a ) + \alpha \mathbb{E}_{\mathbf{s}^{\prime} \given (\mathbf{s}, a)}
      \{ Q(\mathbf{s}^{\prime}, \mu(\mathbf{s}^{\prime})) \}\,, \label{Bellman.standard.mu} \\
    (T^{\diamond} Q)(\mathbf{s}, a)
    & \coloneqq g( \mathbf{s}, a ) + \alpha \mathbb{E}_{\mathbf{s}^{\prime} \given (\mathbf{s}, a)}
      \{ \min_{a^{\prime}\in \mathfrak{A} }Q(\mathbf{s}^{\prime}, a^{\prime})
      \}\,, \label{Bellman.standard}
  \end{align}
\end{subequations}
where $\mathbb{E}_{\mathbf{s}^{\prime} \given (\mathbf{s}, a)}\{\cdot\}$ stands for the conditional
expectation operator with respect to $\mathbf{s}^{\prime}$ conditioned on $(\mathbf{s}, a)$, and
$\alpha\in [0,1)$ is the discount factor. In the case where $Q$ is considered an element of the
Banach space $\mathcal{L}_{\infty}$ of all (essentially) bounded
functions~\cite{Bartle.book:95}, equipped with the norm $\norm{}_{\infty}$, then it can be
shown that the mappings in \eqref{Bellman.maps.standard} are
contractions~\cite{bertsekas2019reinforcement}, and according to the Banach-Picard
theorem~\cite{HB.PLC.book}, they possess \textit{unique}\/ fixed points
$Q^{\diamond}_{\mu}, Q^{\diamond}$, i.e., points which solve the Bellman equations
$T_{\mu}^{\diamond} Q^{\diamond}_{\mu} = Q^{\diamond}_{\mu}$ and
$T^{\diamond} Q^{\diamond} = Q^{\diamond}$, and which characterize ``optimal'' long-term
losses~\cite{bertsekas2019reinforcement}.

Nevertheless, in most cases of practical interest, there is not sufficient information on the
conditional PDF to compute $\mathbb{E}_{\mathbf{s}^{\prime} \given (\mathbf{s},
  a)}\{\cdot\}$. Motivated by this fact, this study proposes approximations of the Bellman
mappings in \eqref{Bellman.maps.standard} by assuming that losses $g$ and $Q$ belong to an RKHS
$\mathcal{H}$, i.e., a Hilbert space with inner product $\innerp{\cdot}{\cdot}_{\mathcal{H}}$,
norm $\norm{\cdot}_{\mathcal{H}} \coloneqq \innerp{\cdot}{\cdot}_{\mathcal{H}}^{1/2}$, and a
reproducing kernel $\kappa(\cdot, \cdot): \mathfrak{Z}\times \mathfrak{Z} \to \Real$, such that
$\kappa(\mathbf{z}, \cdot)\in \mathcal{H}$, $\forall \mathbf{z}\in \mathfrak{Z}$, and the
reproducing property holds true:
$Q(\mathbf{z}) = \innerp{Q}{\kappa(\mathbf{z}, \cdot)}_{\mathcal{H}}$,
$\forall Q\in \mathcal{H}$, $\forall \mathbf{z}\in \mathfrak{Z}$. Space $\mathcal{H}$ may be
infinite dimensional; e.g., $\kappa(\cdot, \cdot)$ is a Gaussian kernel~\cite{aronszajn1950,
  scholkopf2002learning}. For compact notations, let
$\varphi(\mathbf{z}) \coloneqq \kappa(\mathbf{z},\cdot)$, and
$Q^{\intercal} Q^{\prime} \coloneqq \innerp{Q}{Q^{\prime}}_{\mathcal{H}}$.

Hereafter, losses $g, Q$ are assumed to belong to $\mathcal{H}$. The proposed Bellman mappings
$T_{\mu}: \mathcal{H} \to \mathcal{H}: Q\mapsto T_{\mu}Q$ and
$T: \mathcal{H} \to \mathcal{H}: Q\mapsto TQ$ are defined as:
\begin{subequations}\label{Bellman.maps.new}
  \begin{align}
    T_{\mu} Q & \coloneqq g + \alpha \sum\nolimits_{j=1}^{N_{\textnormal{av}}}
                Q (\mathbf{s}^{\textnormal{av}}_j, \mu(\mathbf{s}^{\textnormal{av}}_j) ) \cdot \psi_j
                \,, \label{Bellman.new.mu}\\
    TQ & \coloneqq g + \alpha \sum\nolimits_{j=1}^{N_{\textnormal{av}}} \inf\nolimits_{a_j \in
         \mathfrak{A}} Q (\mathbf{s}^{\textnormal{av}}_j, a_j) \cdot \psi_j
         \,, \label{Bellman.new}
  \end{align}
\end{subequations}
where $\{\psi_j\}_{j=1}^{N_{\textnormal{av}}} \subset \mathcal{H}$, for a user-defined positive
integer $N_{\textnormal{av}}$, and
$\{ \mathbf{s}^{\textnormal{av}}_j \}_{j=1}^{N_{\textnormal{av}}}$ are state vectors chosen by
the user for the summations in \eqref{Bellman.maps.new} to approximate the conditional
expectations in \eqref{Bellman.maps.standard}. For example,
$\{ \mathbf{s}^{\textnormal{av}}_j \}_{j=1}^{N_{\textnormal{av}}}$ may be samples drawn from a
Gaussian PDF centered at a state of interest (the current state $\mathbf{s}_n$ in
\cref{sec:tests}). For notational convenience, let
$\bm{\Psi} \coloneqq [\psi_1, \ldots, \psi_{N_{\textnormal{av}}} ]$, and its
$N_{\textnormal{av}} \times N_{\textnormal{av}}$ kernel matrix
$\mathbf{K}_{\Psi} \coloneqq \bm{\Psi}^{\intercal} \bm{\Psi}$ whose $(j, j^{\prime})$ entry is
equal to $\innerp{\psi_j}{\psi_{j^{\prime}}}_{\mathcal{H}}$.

Consider a subset
$\mathfrak{B} \coloneqq \{ \mathbf{z}^{\textnormal{b}}_i \}^{N_{\textnormal{b}}}_{i=1} \subset
\mathfrak{Z}$, for some $N_{\textnormal{b}} \in \IntegerPP$. Define also
$\bm{\Phi}^{\textnormal{b}} \coloneqq [\varphi(\mathbf{z}^{\textnormal{b}}_1), \ldots,
\varphi(\mathbf{z}^{\textnormal{b}}_{N_{\textnormal{b}}}) ]$,
$\bm{\Phi}_{\mu}^{\textnormal{av}} \coloneqq [\varphi^{\textnormal{av}}_{\mu,1}, \ldots,
\varphi^{\textnormal{av}}_{\mu,N_{\textnormal{av}}}]$, where
$\varphi^{\textnormal{av}}_{\mu, j} \coloneqq \varphi( \mathbf{s}_j^{\textnormal{av}},
\mu(\mathbf{s}_j^{\textnormal{av}}) )$, and let
$\mathbf{K}^{\textnormal{av,b}}_{\mu} \coloneqq
{\bm{\Phi}}^{\textnormal{av}}_{\mu}{}^{\intercal} \bm{\Phi}^{\textnormal{b}}$. Let also the
kernel matrices
$\mathbf{K}_{\textnormal{b}} \coloneqq {\bm{\Phi}^{\textnormal{b}}}^{\intercal}
\bm{\Phi}^{\textnormal{b}}$, and
$\mathbf{K}^{\textnormal{av}}_{\mu} \coloneqq \bm{\Phi}_{\mu}^{\textnormal{av}}{}^{\intercal}
\bm{\Phi}_{\mu}^{\textnormal{av}}$. In the case vectors in $\mathfrak{B}$ are linearly
independent, $\mathbf{K}_{\textnormal{b}}$ is positive definite. Moreover, consider an
$N_{\textnormal{b}}\times N_{\textnormal{av}}$ matrix $\bm{\Upsilon}$. Define then mappings
$T_{\mu}^{\sharp}, T^{\sharp}: \Real^{N_{\textnormal{b}}} \to \Real^{N_{\textnormal{b}}}$ as
follows: $\forall \bm{\xi}\in \Real^{N_{\textnormal{b}}}$,
\begin{subequations}\label{Bellman.maps.sharp}
  \begin{align}
    T_{\mu}^{\sharp}\bm{\xi}
    & \coloneqq \bm{\eta} + \alpha \bm{\Upsilon} \mathbf{K}^{\textnormal{av,b}}_{\mu}
      \bm{\xi} \,, \label{Bellman.sharp.mu} \\
    T^{\sharp}\bm{\xi}
    & \coloneqq \bm{\eta} + \alpha \bm{\Upsilon} \inf\nolimits_{\mu \in
      \mathcal{M}} \mathbf{K}^{\textnormal{av,b}}_{\mu} \bm{\xi} \,, \label{Bellman.sharp}
  \end{align}
\end{subequations}
It can be verified that the fixed-point set of $T_{\mu}^{\sharp}$ satisfies:
\begin{align}
  \Fix T_{\mu}^{\sharp}
  & \coloneqq \{ \bm{\xi}\in\Real^{ N_{\textnormal{b}} } \given T_{\mu}^{\sharp} \bm{\xi}
    = \bm{\xi}\} \notag\\
  & = \{ \bm{\xi}\in\Real^{ N_{\textnormal{b}} } \given (
  \mathbf{I}_{ N_{\textnormal{b}} } - \alpha \bm{\Upsilon} \mathbf{K}^{\textnormal{av,b}}_{\mu}
  ) \bm{\xi} = \bm{\eta} \} \,, \label{Fix.T.sharp.mu}
\end{align}
where $\mathbf{I}_{ N_{\textnormal{b}} }$ stands for the
$N_{\textnormal{b}} \times N_{\textnormal{b}}$ identity matrix. Since $\Fix T_{\mu}^{\sharp}$
may be empty, \cref{algo:step:update.xi} of \cref{algo:rl-lmp} defines a non-empty affine set
as the superset of \eqref{Fix.T.sharp.mu}. More details on \eqref{Fix.T.sharp.mu} are omitted
due to lack of space. Note here that \cite{Yuki:ICASSP23} focuses on \eqref{Bellman.new.mu} and
follows a different route by defining a potentially infinite-dimensional hyperplane as the
superset of the fixed-point of \eqref{Bellman.new.mu}.

\begin{algorithm}[t]
  \begin{algorithmic}[1]
    \renewcommand{\algorithmicindent}{1em}

    \State{Arbitrarily initialize $\mathfrak{B}_0\subset \mathfrak{Z}$, with
      $N_{\textnormal{b}}[0] = \lvert \mathfrak{B}_0\rvert$, $\bm{\xi}_0 \in
      \mathbb{R}^{N_{\textnormal{b}}[0]}$, $\bm{\Phi}^{\textnormal{b}}_0$, $Q_0$,
      $\mu_0\in\mathcal{M}$, and a $\bm{\theta}_0\in \Real^L$.}

    \While{$n \in \mathbb{N}$}\label{line:iter}

      \State{Data $(\mathbf{x}_n, y_n)$ become available. Let $\mathbf{s}_n = (\mathbf{x}_n,
        y_n, \bm{\theta}_n)$.}

      \State{Define $\{\mathbf{s}_j^{\textnormal{av}}[n]\}_{j=1}^{N_{\textnormal{av}}[n]}$.}

      \State{\textbf{Policy improvement:} Update $\mu_n(\mathbf{s}_n)$ by \eqref{rollout}.}

      \State{Update $\bm{\theta}_{n+1}$ by \eqref{LMP}, where $p \coloneqq
        \mu_n(\mathbf{s}_n)$.}

      \State{Update $\mathfrak{B}_n$, $\bm{\Phi}^{\textnormal{b}}_n$, and identify
        $\bm{\Upsilon}_n$ and $\mathbf{K}^{\textnormal{av,b}}_{\mu_n}$ as in \cref{sec:algo}.}

      \State{Update $\bm{\eta}_n$ by \eqref{choices.eta}.}

      \State{Compute
        \[
          \bm{\xi}_{n+1} \coloneqq \arg\min_{\bm{\xi}}
          \norm*{ (\mathbf{I}_{N_{\textnormal{b}}[n]} - \alpha \bm{\Upsilon}_n
            \mathbf{K}^{\textnormal{av,b}}_{\mu_n} )\bm{\xi} - \bm{\eta}_n}^2\,.
        \]
      }\label{algo:step:update.xi}

      \State{\textbf{Policy evaluation:} Update $Q_{n+1} \coloneqq\bm{\Phi}^{\textnormal{b}}_n
        \bm{\xi}_{n+1}$.}

      \State{Increase $n$ by one, and go to Line \ref{line:iter}.}

    \EndWhile
  \end{algorithmic}

  \caption{Approximate policy-iteration framework.}\label{algo:rl-lmp}

\end{algorithm}

\begin{theorem}\label{thm:nonexp} Let $\psi_j(\mathbf{z})\geq 0$, $\forall \mathbf{z}\in
  \mathfrak{Z}$, $\forall j\in \{1, \ldots, N_{\textnormal{av}}\}$.

  \begin{thmlist}

  \item\label{thm:nonexp.T} If
    $\alpha \leq \norm{\mathbf{K}_{\Psi}}^{-1/2} (\sup_{\mu\in\mathcal{M}} \norm{
      \mathbf{K}^{\textnormal{av}}_{\mu}} )^{-1/2}$, then
    $\forall \mu\ \in \mathcal{M}$, the mapping $T_{\mu}$ in \eqref{Bellman.new.mu} is affine
    nonexpansive and $T$ in \eqref{Bellman.new} is nonexpansive within the Hilbert space
    $( \mathcal{H}, \innerp{\cdot}{\cdot}_{\mathcal{H}} )$. Norms
    $\norm{\mathbf{K}_{\Psi}}, \norm{\mathbf{K}^{\textnormal{av}}_{\mu}}$ are the
    spectral norms of $\mathbf{K}_{\Psi}, \mathbf{K}^{\textnormal{av}}_{\mu}$.

  \item\label{thm:nonexp.T.sharp} Let
    $\alpha \leq \norm{\mathbf{K}_{\Psi}}^{-1/2} (\sup_{\mu\in\mathcal{M}} \norm{
      \mathbf{K}^{\textnormal{av}}_{\mu}} )^{-1/2}$. Consider also the case where vectors
    $\{ \varphi(\mathbf{z}^{\textnormal{b}}_i) \}^{N_{\textnormal{b}}}_{i=1}$ are linearly
    independent, and
    $\{ g, \{\psi_j\}_{j=1}^{N_{\textnormal{av}}} \}\subset \mathcal{H}_{\textnormal{b}}
    \coloneqq \linspan ( \{ \varphi(\mathbf{z}^{\textnormal{b}}_i)
    \}^{N_{\textnormal{b}}}_{i=1} )$, that is, there exist
    $\bm{\eta}\in \Real^{N_{\textnormal{b}}}$ and
    $\bm{\Upsilon}\in \Real^{N_{\textnormal{b}} \times N_{\textnormal{av}}}$ such that
    $g = \bm{\Phi}^{\textnormal{b}}\bm{\eta}$ and
    $\bm{\Psi} = \bm{\Phi}^{\textnormal{b}} \bm{\Upsilon}$. Then, for any
    $Q\in \mathcal{H}_{\textnormal{b}}$,
    $T_{\mu} Q = \bm{\Phi}^{\textnormal{b}} T_{\mu}^{\sharp} \bm{\xi}$,
    $\forall\mu \in \mathcal{M}$, and $TQ = \bm{\Phi}^{\textnormal{b}}
    T^{\sharp}\bm{\xi}$. Moreover, $\forall \mu\ \in \mathcal{M}$, the mapping
    $T_{\mu}^{\sharp}$ in \eqref{Bellman.sharp.mu} is affine nonexpansive and $T^{\sharp}$ in
    \eqref{Bellman.sharp} is nonexpansive within the Euclidean space
    $(\mathbb{R}^{N_{\textnormal{b}}}, \innerp{\cdot}{\cdot}_{\mathbf{K}_{\textnormal{b}}} )$,
    where
    $\innerp{ \bm{\xi}}{\bm{\xi'}}_{\mathbf{K}_{\textnormal{b}}} \coloneqq \bm{\xi}^{\intercal}
    \mathbf{K}_{\textnormal{b}}\bm{\xi'}$.

  \end{thmlist}

\end{theorem}

Nonexpansivity for a mapping $T$ in a (Euclidean) Hilbert space
$( \mathcal{H}, \innerp{\cdot}{\cdot}_{\mathcal{H}} )$ means
$\norm{ TQ - TQ^{\prime}}_{\mathcal{H}} \leq \norm{ Q - Q^{\prime}}_{\mathcal{H}}$,
$\forall Q, Q^{\prime}\in \mathcal{H}$~\cite{HB.PLC.book}. Moreover, mapping
$T_{\mu}: \mathcal{H} \to \mathcal{H}$ is affine iff
$T_{\mu}( \lambda Q + (1-\lambda)Q^{\prime}) = \lambda TQ + (1 - \lambda)TQ^{\prime}$,
$\forall Q, Q^{\prime}\in \mathcal{H}$, $\forall\lambda \in \Real$.

Mappings \eqref{Bellman.maps.new} share similarities with the mappings in
\cite{ormoneit2002kernel, Ormoneit:Autom:02, Barreto:NIPS:11, Barreto:NIPS:12,
  Kveton_Theocharous_2013, Kveton_Theocharous_2021}. However, in~\cite{ormoneit2002kernel,
  Ormoneit:Autom:02, Barreto:NIPS:11, Barreto:NIPS:12, Kveton_Theocharous_2013,
  Kveton_Theocharous_2021} as well as in the classical context of
\eqref{Bellman.maps.standard}, Bellman mappings are viewed as contractions on the Banach space
of (essentially) bounded functions with the
$\mathcal{L}_{\infty}$-norm~\cite{bertsekas2019reinforcement}, while no discussion on RKHSs is
reported. Recall that, by definition, Banach spaces are not equipped with inner products. On
the other hand, \cref{thm:nonexp} opens the door to the rich toolbox of nonexpansive mappings
in Hilbert spaces~\cite{HB.PLC.book}, to the reproducing property of the inner product in
RKHSs~\cite{aronszajn1950, scholkopf2002learning}, and also to the familiar Euclidean spaces
with their convenient norms by \eqref{Bellman.maps.sharp} and \cref{thm:nonexp.T.sharp}.

Due to lack of space, a detailed discussion on ways to choose/design
$\{\psi_j\}_{j=1}^{N_{\textnormal{av}}}$,
$\{ \mathbf{s}^{\textnormal{av}}_j \}_{j=1}^{N_{\textnormal{av}}}$, and
$\{\mathbf{z}^{\textnormal{b}}_i \}^{N_{\textnormal{b}}}_{i=1}$ will be reported in the journal
version of the paper. A way to choose $\{\psi_j\}_{j=1}^{N_{\textnormal{av}}}$ is reported
in~\cite{Yuki:ICASSP23}.

\section{Approximate Policy-Iteration Framework}\label{sec:algo}

Due to the approximate Bellman mappings offered in \eqref{Bellman.maps.new} and
\eqref{Bellman.maps.sharp}, the proposed \textit{approximate}\/ PI framework for the problem at
hand is summarized in \cref{algo:rl-lmp}. The framework operates sequentially, with its
iteration index $n$ coinciding with the time index of the streaming data
$(\mathbf{x}_n, y_n)_{n\in \IntegerP}$ which are generated according to the discussion
surrounding \eqref{LMP}. To this end, the arguments of \cref{sec:nonexp.Bellman} will be
adapted to include hereafter the extra time dimension $n$, which will be indicated by the
super-/sub-scripts $[n]$, $(n)$ or $n$ in notations.

With
$\mathfrak{B}[n-1] = \{\mathbf{z}^{\textnormal{b}}_i[n-1] \}^{ N_{\textnormal{b}}[n-1] }_{i=1}$
available at time $n$, and with
$\mathcal{H}_{\textnormal{b}}[n-1] = \linspan ( \{ \varphi(\mathbf{z}^{\textnormal{b}}_i[n-1])
\}^{ N_{\textnormal{b}}[n-1] }_{i=1} )$, ALD~\cite{engel2004krls} is used here to let
\cref{algo:rl-lmp} decide whether $\mathfrak{B}[n-1]$ is updated to $\mathfrak{B}[n]$ by
including a new basis vector or not. To reduce as much as possible the unpleasant effects of
the ``curse of dimensionality,'' $\mathbf{s}_n = ( \mathbf{x}_n, y_n, \bm{\theta}_n )$ is
projected onto the compact unit ball to define
$\mathbf{s}_n^{\|} \coloneqq \mathbf{s}_n/ \norm{\mathbf{s}_n}$, and the potentially new basis
vector
$\varphi( \mathbf{z}^{\textnormal{b}}_{N_{\textnormal{b}}[n]}[n] ) \coloneqq
\kappa(\mathbf{z}^{\textnormal{b}}_{N_{\textnormal{b}}[n]}[n], \cdot)$ of
$\mathcal{H}_{\textnormal{b}}[n]$, with
$\mathbf{z}^{\textnormal{b}}_{N_{\textnormal{b}}[n]}[n] \coloneqq (\mathbf{s}_n^{\|},
\mu_n(\mathbf{s}_n) )$, is tested against $\mathcal{H}_{\textnormal{b}}[n-1]$ according to
ALD~\cite{engel2004krls}. Note also that ALD ensures that vectors in
$\mathcal{H}_{\textnormal{b}}[n]$ are linearly independent, as needed in
\cref{thm:nonexp.T.sharp}. ALD is also utilized upon the arrival of
$\{ \mathbf{s}^{\textnormal{av}}_j[n] \}^{N_\textnormal{av}[n]}_{j=1}$ to check whether any of
these vectors enter the basis or not.

To utilize \eqref{Bellman.sharp.mu}, an $N_{\textnormal{b}}[n] \times 1$ vector $\bm{\eta}_n$
needs to be identified. According to \cref{thm:nonexp.T.sharp}, such an $\bm{\eta}_n$
identifies a vector $\tilde{g}_n \coloneqq \bm{\Phi}^{\textnormal{b}}_n \bm{\eta}_n$ in
$\mathcal{H}_{\textnormal{b}}[n]$. More specifically, $\bm{\eta}_n$ is chosen such that
$[ \tilde{g}_n( \mathbf{z}_n ) - g( \mathbf{z}_n )]^2$ is minimized, where
$\mathbf{z}_n = ( \mathbf{s}_n, \mu_n(\mathbf{s}_n) )$, and the value $g( \mathbf{z}_n )$ is
provided by \eqref{one.step.loss}. Here is where the reproducing property of the inner product
in $\mathcal{H}$ becomes handy:
$\tilde{g}_n( \mathbf{z}_n ) = \innerp{\tilde{g}_n} {\varphi(\mathbf{z}_n)}_{\mathcal{H}} =
\innerp{\bm{\Phi}^{\textnormal{b}}_n \bm{\eta}_n} {\varphi(\mathbf{z}_n)}_{\mathcal{H}} =
\mathbf{k}^{\intercal}(\mathbf{z}_n) \bm{\eta}_n$, where
$\mathbf{k}(\mathbf{z}_n) \coloneqq \bm{\Phi}^{\textnormal{b}}_n{}^{\intercal} \varphi(\mathbf{z}_n)
= [ \kappa(\mathbf{z}^{\textnormal{b}}_1[n], \mathbf{z}_n), \ldots,
\kappa(\mathbf{z}^{\textnormal{b}}_{N_{\textnormal{b}}[n]}[n], \mathbf{z}_n) ]^{\intercal}$. Hence,
$\bm{\eta}_n$ can be chosen as:
\begin{align}
  \bm{\eta}_n \in
  {} & {} \arg\min\nolimits_{\bm{\eta} \in \Real^{N_{\textnormal{b}}[n]}} [
    \mathbf{k}^{\intercal}(\mathbf{z}_n) \bm{\eta} - g(\mathbf{z}_n) ]^2 \notag\\
  {} & {} = \{ \bm{\eta} \in \Real^{N_{\textnormal{b}}[n]} \given  \mathbf{k}^{\intercal}(\mathbf{z}_n)
    \bm{\eta} = g(\mathbf{z}_n)\} \,. \label{eq:hyperplane}
\end{align}
To this end, the following options are provided to choose from:
\begin{subnumcases}{ \bm{\eta}_n \coloneqq \label{choices.eta}}
  \frac{g(\mathbf{z}_n)}{ \mathbf{k}^{\intercal}(\mathbf{z}_n) \mathbf{K}_{\textnormal{b}}[n]
    \mathbf{k}(\mathbf{z}_n) } \mathbf{k}(\mathbf{z}_n) \,, \label{min.norm}\\
  \bm{\eta}_{n-1} + \frac{ g(\mathbf{z}_n) - \bm{\eta}_{n-1}^{\intercal} \mathbf{K}_{\textnormal{b}}[n]
    \mathbf{k}(\mathbf{z}_n) }{ \mathbf{k}^{\intercal}(\mathbf{z}_n) \mathbf{K}_{\textnormal{b}}[n]
    \mathbf{k}(\mathbf{z}_n) } \mathbf{k}(\mathbf{z}_n) \,, \label{Proj.previous.eta}
\end{subnumcases}
where \eqref{min.norm} offers the minimum-norm solution of \eqref{eq:hyperplane}, while
\eqref{Proj.previous.eta} computes the (metric) projection~\cite{HB.PLC.book} of
$\bm{\eta}_{n-1}$ onto the hyperplane in \eqref{eq:hyperplane}, after zero-padding the
$N_{\textnormal{b}}[n-1]\times 1$ vector $\bm{\eta}_{n-1}$ to meet the length
$N_{\textnormal{b}}[n]$, in case $N_{\textnormal{b}}[n] \geq N_{\textnormal{b}}[n-1] +
1$. Following \cref{thm:nonexp.T.sharp}, the inner product
$\innerp{\cdot}{\cdot}_{\mathbf{K}_{\textnormal{b}}[n]}$ was utilized in \eqref{choices.eta}.

To identify $\bm{\Upsilon}_n$, needed for the computations in \cref{algo:step:update.xi} of
\cref{algo:rl-lmp}, \cref{thm:nonexp.T.sharp} is followed where $\bm{\Upsilon}_n$ is computed
via $\bm{\Psi}_n$. Although $\bm{\Psi}_n$ can be chosen in many ways such that
$\{\psi_j[n]\}_{j=1}^{N_{\textnormal{av}}[n]} \subset \mathfrak{B}[n]$,
$\{\psi_j[n]\}_{j=1}^{N_{\textnormal{av}}[n]}$ are selected in \cref{sec:tests} simply by
taking the first $N_{\textnormal{av}}[n]$ vectors from $\mathfrak{B}[n]$, under the assumption
that $N_{\textnormal{av}}[n] \leq N_{\textnormal{b}}[n]$. Hence, according to
\cref{thm:nonexp.T.sharp},
$\bm{\Upsilon}_n = [ \mathbf{I}_{N_{\textnormal{av}}[n]}, \mathbf{0}_{ N_{\textnormal{av}}[n]
  \times (N_{\textnormal{b}}[n] - N_{\textnormal{av}}[n]) } ]^{\intercal}$ in
\cref{sec:tests}. With regards to
$\varphi^{\textnormal{av}}_{\mu, j}[n] \coloneqq \varphi( \mathbf{s}_j^{\textnormal{av}}[n],
\mu_n(\mathbf{s}_j^{\textnormal{av}}[n]) )$, needed in the computation of
$\mathbf{K}^{\textnormal{av,b}}_{\mu_n}$, since $\mathbf{s}_j^{\textnormal{av}}[n]$ is drawn
from a Gaussian PDF centered at $\mathbf{s}_n$, $\mu_n(\mathbf{s}_j^{\textnormal{av}}[n])$ is
set equal to $\mu_n(\mathbf{s}_n)$ in \cref{sec:tests}.

Typically, the greedy rule
$\mu_{n} \left(\mathbf{s}_n \right) = \arg \min_{a\in \mathfrak{A}} Q_n(\mathbf{s}_n,a)$ is
used for policy improvement~\cite{bertsekas2019reinforcement, lagoudakis2003lspi,
  xu2007klspi}. However, it has been observed that this greedy rule may lead into instabilities
of the RL agent's behavior and hinder its cognition about the surrounding environment
\cite{bertsekas2019reinforcement}. The $\varepsilon$-greedy strategy
\cite{bertsekas2019reinforcement}, a variant of the greedy one, has not performed well in the
numerical tests of \cref{sec:tests}. To address this potential drawback of greedy strategies,
the popular rollout methodology~\cite{bertsekas2019reinforcement} is employed in
\cref{algo:rl-lmp}. With the user-defined $M\in \IntegerPP$ denoting the number of steps in
rollout, the following rule is employed for policy improvement:
\begin{align}
  \mu_{n} (\mathbf{s}_n) = \arg \min_{a \in \mathfrak{A}} \Big[
  & g(\mathbf{s}_n, a) +
    \sum\nolimits_{m=1}^{M-1} \alpha^{m} g(\mathbf{s}_{n+m}^{\triangleright}, \mu^{\triangleright}(
    \mathbf{s}_{n+m}^{\triangleright})) \notag \\
  & + \alpha^{M} Q_n(\mathbf{s}_{n+M}^{\triangleright}, \mu^{\triangleright}(
    \mathbf{s}_{n+M}^{\triangleright})) \Big] \,, \label{rollout}
\end{align}
where $\mu^{\triangleright} \in\mathcal{M}$ is called the \textit{heuristic}\/ (stationary)
policy, which could be chosen to be random or $\mu_{n-1}$, and
$\{\mathbf{s}_{n+m}^{\triangleright}\}_{m=1}^M$, denote possible successor states of
$\mathbf{s}_n$ under action $a$ and policy $\mu^{\triangleright}$. A large $M$ (long
trajectory) may inflict long computational times and diminish the effect of $Q_n$ on policy
improvement through $\alpha^M$ in \eqref{rollout}; usually,
$\alpha \leq 1$~\cite{bertsekas2019reinforcement}. Note here that experience
replay~\cite{ExperienceReplay} and not rollout was employed in~\cite{Yuki:ICASSP23}.

\section{Numerical Tests}\label{sec:tests}

\begin{figure}[t]
    \centering

    \subfloat[``Gauss-like'' outliers]{ \includegraphics[ width = .225\textwidth
      ]{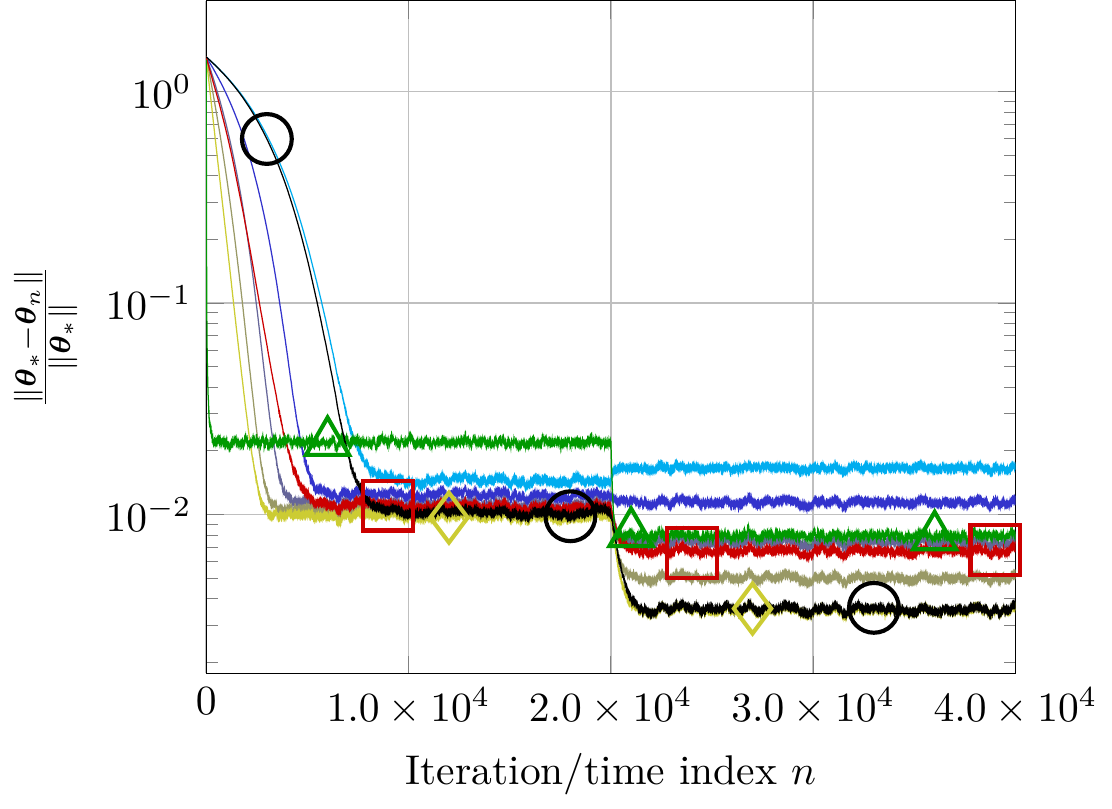}\label{fig:gauss-like}}
    \subfloat[``Cauchy-like'' outliers]{ \includegraphics[ width =.225\textwidth
      ]{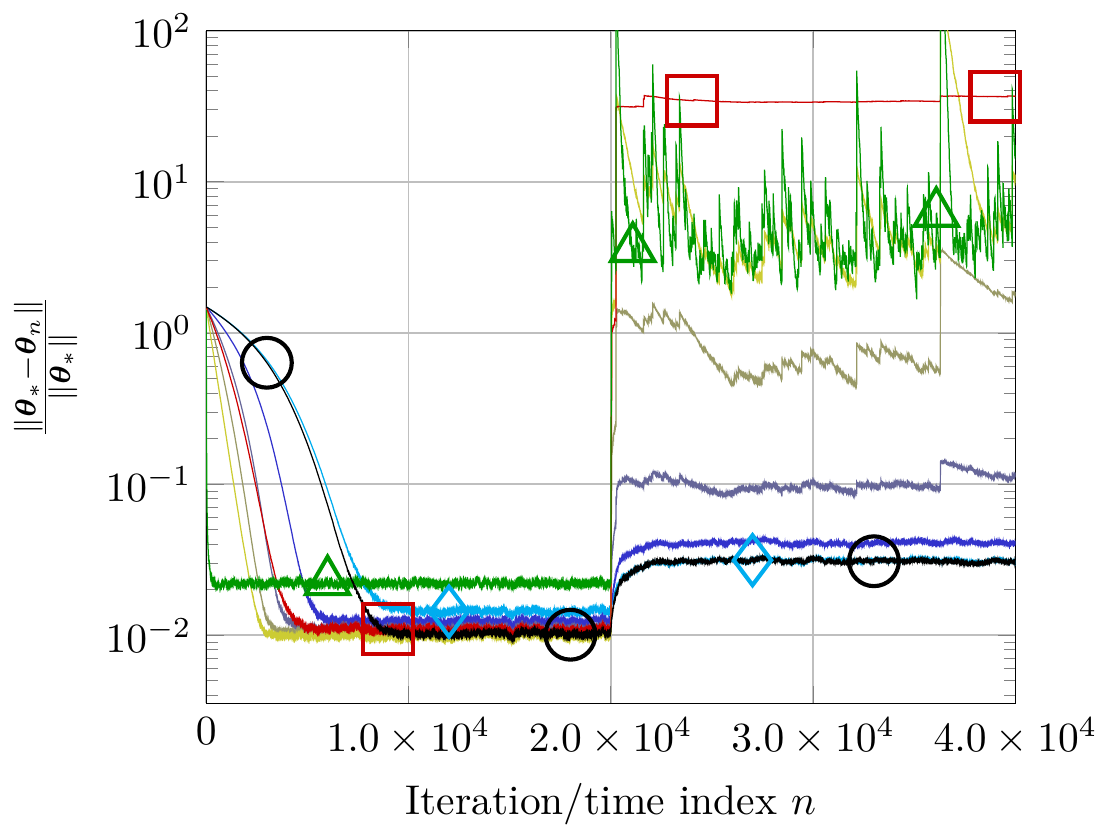}\label{fig:cauchy-like}}

    \caption[]{Outliers appear only at $n\geq 2\times 10^4$. Curve markers: \cref{algo:rl-lmp}:
      \tikz{ \node[ mark size = 3pt, black, line width = 1pt ]{\pgfuseplotmark{o}};}, KLSPI
      \cite{xu2007klspi}: \tikz{\node[mark size = 3pt, red!80!black, line width =
        1pt]{\pgfuseplotmark{square}};}, CAC-RL$p$ \cite{vazquez2012}: \tikz{ \node[mark
        size=3.5pt, green!60!black, line width=1pt]{\pgfuseplotmark{triangle}};}, optimal
      $p$-norm criterion, $2$-norm: \tikz{\node[mark size=4pt, blue!20!yellow, line
        width=1pt]{\pgfuseplotmark{diamond}};} for \cref{fig:gauss-like} and $1$-norm:
      \tikz{\node[mark size=4pt, cyan, line width=1pt]{\pgfuseplotmark{diamond}};} for
      \cref{fig:cauchy-like}. }\label{fig:results}

\end{figure}

\begin{figure}[t]
    \centering

    \subfloat[``Gauss-like'' outliers]{ \includegraphics[ width =.225\textwidth
      ]{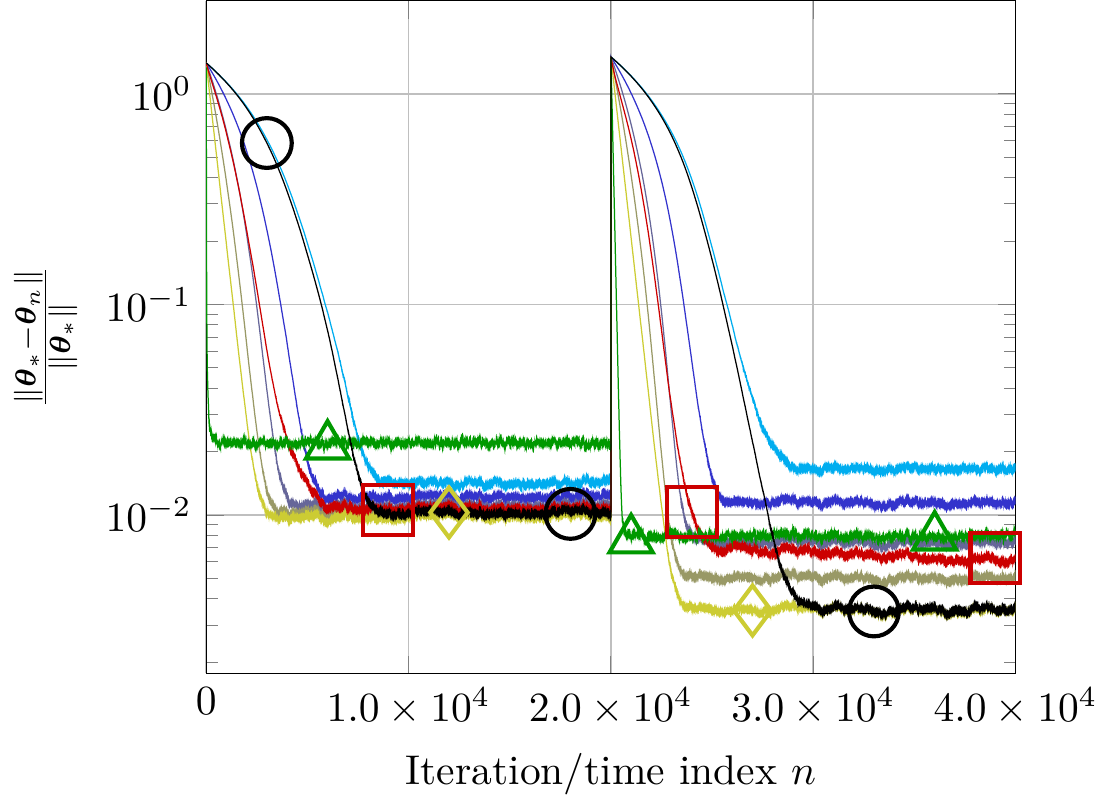}\label{fig:gauss-like-theta-change}}
    \subfloat[``Cauchy-like'' outliers]{ \includegraphics[ width =.225\textwidth
      ]{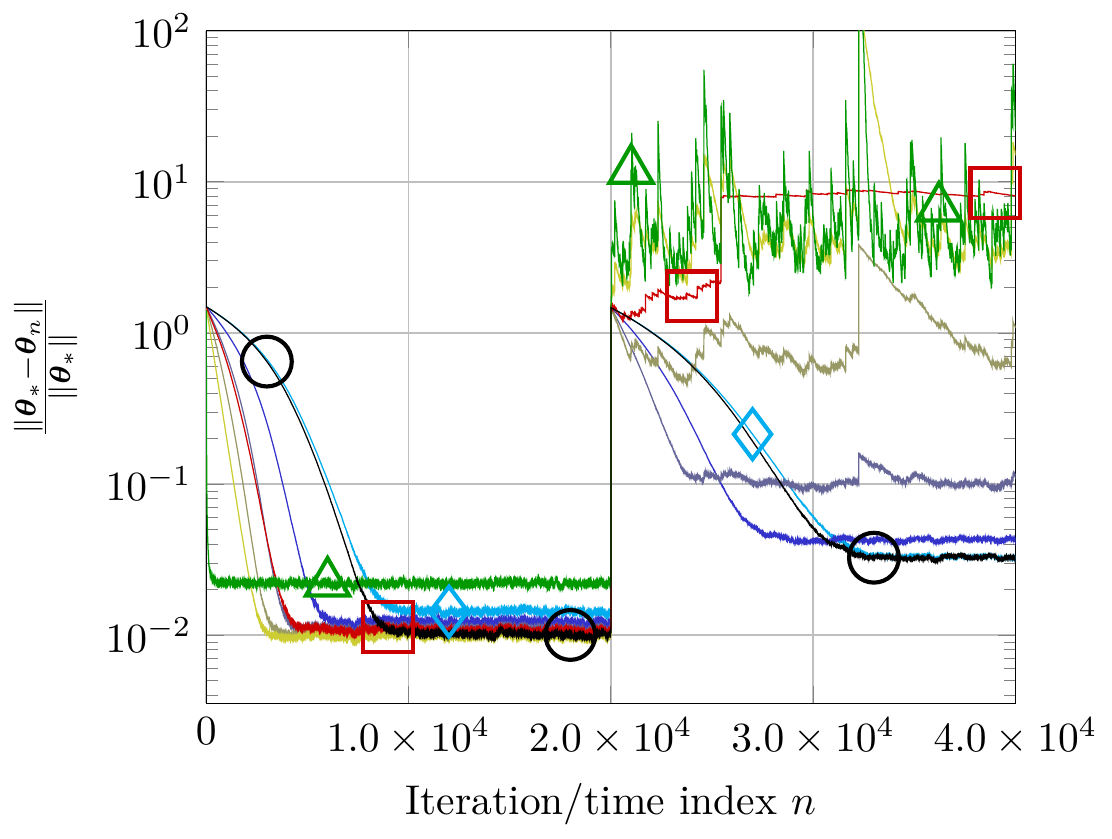}\label{fig:cauchy-like-theta-change}}

    \caption[]{System $\bm{\theta}_*$ changes randomly at $n = 2\times 10^4$, and outliers
      appear only at $n\geq 2\times 10^4$. Curve markers follow those of
    \cref{fig:results}.}\label{fig:results-2}
\end{figure}

\sloppy The proposed PI framework is compared numerically against
\begin{enumerate*}[label=\textbf{(\roman*)}]
\item \eqref{LMP}, where $p$ is fixed throughout iterations with
  $p \in \mathfrak{A} \coloneqq \{ 1, 1.25, 1.5, 1.75, 2 \}$,
\item CAC-RL$p$~\cite{vazquez2012}, which uses a combination of $p$-norm adaptive filters with
  different forgetting factors, and
\item \cite{xu2007klspi}, which is based on the kernel-based least-squares temporal-difference
  methodology with no rollout employed.
\end{enumerate*}
Performance is measured by the normalized deviation
$\lVert \bm{\theta}_* - \bm{\theta}_n \rVert/\lVert \bm{\theta}_* \rVert$ vs.\ time index
$n$. Multiple ($100$) independent tests were performed, with their uniformly averaged results
reported in \Cref{fig:results,fig:results-2}. The software code was written in
Julia~\cite{bezanson2017julia}, with $\alpha$-stable outliers generated
by~\cite{miotto2016pylevy}.

The length $L$ of the system $\bm{\theta}_*$ is set equal to $10$ in the data-generation model
of \cref{sec:intro}. The entries of $\bm{\theta}_*$ and $\mathbf{x}_n$, $\forall n$, are
generated by independent and identically distributed (IID) normal RVs. In the data-generation
model of \cref{sec:intro}, RV $o_n$ describes both outlier and noise scenarios. In the case
where no outliers appear in the data-generation model, $o_n$ obeys the Gaussian PDF, and
whenever outliers appear, $o_n$ follows the $\alpha$-stable PDF~\cite{miotto2016pylevy}. In all
figures, no outliers appear for time instances $n < 2\times 10^4$ and only Gaussian noise, with
$\textnormal{SNR} = 20\textnormal{dB}$, corrupts the data. Outliers appear for
$n \geq 2\times 10^4$, following the $\alpha$-stable PDF~\cite{shao1993signal}. Note
that when the ``stability'' parameter of the $\alpha$-stable PDF takes the values of
$2$, then the PDF boils down to the Gaussian one. Two types of $\alpha$-stable
outliers are examined:
\begin{enumerate*}[label=\textbf{(\roman*)}]
\item ``Gaussian-like'' ones, with parameters $\textnormal{"stability"} = 1.95$,
  $\textnormal{"skewness"} = 0.5$, $\textnormal{"location"} = 0.5$,
  $\textnormal{"scale"} = 10^{-2}$, which make the tails of the PDF slightly heavier
  than those of the Gaussian PDF~\cite{miotto2016pylevy}; and
\item ``Cauchy-like'' ones, with parameters $\textnormal{"stability"} = 1$,
  $\textnormal{"skewness"} = 0.5$, $\textnormal{"location"} = 0.5$ and
  $\textnormal{"scale"} = 1$, which make the tails of the PDF rather
  heavy~\cite{miotto2016pylevy}.
\end{enumerate*}

Number $N_{\textnormal{av}}[n] = 10$, $\forall n$, $M=2$ in \eqref{rollout}, while
\eqref{min.norm} is employed. The learning rate $\rho$ of LMP in \eqref{LMP} is set equal to
$10^{-3}$. A homogeneous polynomial kernel of degree $2$ is used for both \cref{algo:rl-lmp}
and KLSPI \cite{xu2007klspi}. Note that this kernel satisfies the condition stated in the first
line of \cref{thm:nonexp}. Results on other kernels will be reported elsewhere. The ALD
criterion is employed with parameter $\delta_{\textnormal{ALD}} = \sin (40\pi/180)$. The
parameters of CAC-RL$p$ are set as $p_{\textnormal{RLP}} = 1.4$,
$\beta_{\textnormal{RLP}} = 0.9$ and $\lambda_{\textnormal{RLP}} = 0.99$.

All tests show that \cref{algo:rl-lmp} succeeds \textit{always}\/ in identifying the
``optimal'' $p$-norm, regardless of the PDF of the outliers. Even if outliers are absent from
the data-generation model and only Gaussian noise corrupts the data, \cref{algo:rl-lmp} chooses
the $2$-norm, which is well-known to be optimal for Gaussian noise. In contrast, KLSPI
identifies fast the $2$-norm in the case where only Gaussian noise appears, but performs poorly
when Cauchy-like outliers appear. CAC-RL$p$~\cite{vazquez2012} converges fast in cases where
only Gaussian noise appears, with sub-optimal performance in terms of the normalized deviation,
but diverges in the case where Cauchy-like outliers corrupt the data. Due to space limitations,
further numerical tests will be reported in the journal version of this manuscript.

\clearpage
\footnotesize
\bibliographystyle{ieeetr}
\bibliography{refs}

\begin{thebibliography}{10}

\bibitem{sayed2011adaptive}
A.~H. Sayed, {\em Adaptive Filters}.
\newblock Wiley, 2011.

\bibitem{Theodoridis.Book:ML}
S.~Theodoridis, {\em Machine Learning---A Bayesian and Optimization
  Perspective}.
\newblock Elsevier, 2nd~ed., 2020.

\bibitem{rousseeuw1987}
P.~J. Rousseeuw and A.~Leroy, {\em Robust Regression and Outlier Detection}.
\newblock Wiley, 1987.

\bibitem{shao1993signal}
M.~Shao and C.~L. Nikias, ``Signal processing with fractional lower order
  moments: {S}table processes and their applications,'' {\em Proceedings of the
  IEEE}, vol.~81, no.~7, pp.~986--1010, 1993.

\bibitem{miotto2016pylevy}
J.~M. Miotto, ``Pylevy.'' \url{https://github.com/josemiotto/pylevy}, 2020.

\bibitem{pei1994p-power}
S.-C. Pei and C.-C. Tseng, ``Least mean p-power error criterion for adaptive
  {FIR} filter,'' {\em IEEE Journal on Selected Areas in Communications},
  vol.~12, no.~9, pp.~1540--1547, 1994.

\bibitem{xiao1999adaptive}
Y.~Xiao, Y.~Tadokoro, and K.~Shida, ``Adaptive algorithm based on least mean
  p-power error criterion for {F}ourier analysis in additive noise,'' {\em IEEE
  Transactions on Signal Processing}, vol.~47, no.~4, pp.~1172--1181, 1999.

\bibitem{Kuruoglu:02}
E.~E. Kuruo\u{g}lu, ``Nonlinear least $\ell_p$-norm filters for nonlinear
  autoregressive $\alpha$-stable processes,'' {\em Digital Signal Processing},
  vol.~12, no.~1, pp.~119--142, 2002.

\bibitem{vazquez2012}
A.~Navia-Vazquez and J.~Arenas-Garcia, ``Combination of recursive least p-norm
  algorithms for robust adaptive filtering in alpha-stable noise,'' {\em IEEE
  Transactions on Signal Processing}, vol.~60, no.~3, pp.~1478--1482, 2012.

\bibitem{chen2015smoothed}
B.~Chen, L.~Xing, Z.~Wu, J.~Liang, J.~C. Pr\'{\i}ncipe, and N.~Zheng,
  ``Smoothed least mean p-power error criterion for adaptive filtering,'' {\em
  Digital Signal Processing}, vol.~40, pp.~154--163, May 2015.

\bibitem{slavakis2021outlier}
K.~Slavakis and M.~Yukawa, ``Outlier-robust kernel hierarchical-optimization
  {RLS} on a budget with affine constraints,'' in {\em Proc.\ IEEE ICASSP},
  pp.~5335--5339, 2021.

\bibitem{Singh.MCC:09}
A.~Singh and J.~C. Pr\'{i}ncipe, ``Using correntropy as a cost function in
  linear adaptive filters,'' in {\em Proc.\ International Joint Conference on
  Neural Networks}, pp.~2950--2955, 2009.

\bibitem{Gentile:03}
C.~Gentile, ``The robustness of the p-norm algorithms,'' {\em Machine
  Learning}, vol.~53, pp.~265--299, 2003.

\bibitem{bertsekas2019reinforcement}
D.~Bertsekas, {\em Reinforcement Learning and Optimal Control}.
\newblock Athena Scientific, 2019.

\bibitem{ormoneit2002kernel}
D.~Ormoneit and S.~Sen, ``Kernel-based reinforcement learning,'' {\em Machine
  Learning}, vol.~49, pp.~161--178, 2002.

\bibitem{Ormoneit:Autom:02}
D.~Ormoneit and P.~Glynn, ``Kernel-based reinforcement learning in average-cost
  problems,'' {\em IEEE Transactions on Automatic Control}, vol.~47,
  pp.~1624--1636, Oct. 2002.

\bibitem{xu2007klspi}
X.~Xu, D.~Hu, and X.~Lu, ``Kernel-based least squares policy iteration for
  reinforcement learning,'' {\em IEEE Transactions on Neural Networks},
  vol.~18, no.~4, pp.~973--992, 2007.

\bibitem{DDQN}
H.~Van~Hasselt, A.~Guez, and D.~Silver, ``Deep reinforcement learning with
  double {Q}-learning,'' in {\em Proc.\ AAAI conference on Artificial
  Intelligence}, vol.~30, 2016.

\bibitem{Bae:MLSP:11}
J.~Bae, L.~S. Giraldo, P.~Chhatbar, J.~Francis, J.~Sanchez, and
  J.~Pr\'{\i}ncipe, ``Stochastic kernel temporal difference for reinforcement
  learning,'' in {\em Proc.\ IEEE MLSP}, pp.~1--6, 2011.

\bibitem{Barreto:NIPS:11}
A.~Barreto, D.~Precup, and J.~Pineau, ``Reinforcement learning using
  kernel-based stochastic factorization,'' in {\em Proc.\ NIPS}, vol.~24, 2011.

\bibitem{Barreto:NIPS:12}
A.~Barreto, D.~Precup, and J.~Pineau, ``On-line reinforcement learning using
  incremental kernel-based stochastic factorization,'' in {\em Proc.\ NIPS},
  vol.~25, 2012.

\bibitem{Kveton_Theocharous_2013}
B.~Kveton and G.~Theocharous, ``Structured kernel-based reinforcement
  learning,'' in {\em Proc.\ AAAI Conference on Artificial Intelligence},
  vol.~27, pp.~569--575, June 2013.

\bibitem{OnlineBRloss:16}
W.~Sun and J.~A. Bagnell, ``Online {B}ellman residual and temporal difference
  algorithms with predictive error guarantees,'' in {\em Proc.\ International
  Joint Conference on Artificial Intelligence}, pp.~4213--4217, 2016.

\bibitem{RegularizedPI:16}
A.-M. Farahmand, M.~Ghavamzadeh, C.~Szepesv{\'a}ri, and S.~Mannor,
  ``Regularized policy iteration with nonparametric function spaces,'' {\em J.\
  Machine Learning Research}, vol.~17, no.~1, pp.~4809--4874, 2016.

\bibitem{Kveton_Theocharous_2021}
B.~Kveton and G.~Theocharous, ``Kernel-based reinforcement learning on
  representative states,'' in {\em Proc.\ AAAI Conference on Artificial
  Intelligence}, vol.~26, pp.~977--983, Sept. 2021.

\bibitem{Wang_Principe:SPM:21}
Y.~Wang and J.~C. Pr\'{\i}ncipe, ``Reinforcement learning in reproducing kernel
  {H}ilbert spaces,'' {\em IEEE Signal Processing Magazine}, vol.~38, no.~4,
  pp.~34--45, 2021.

\bibitem{bellman2003dp}
R.~E. Bellman, {\em Dynamic Programming}.
\newblock Dover Publications, 2003.

\bibitem{aronszajn1950}
N.~Aronszajn, ``Theory of reproducing kernels,'' {\em Transactions of the
  American Mathematical Society}, vol.~68, pp.~337--404, 1950.

\bibitem{scholkopf2002learning}
B.~Sch\"{o}lkopf and A.~J. Smola, {\em Learning with Kernels: Support Vector
  Machines, Regularization, Optimization, and Beyond}.
\newblock Adaptive computation and machine learning, MIT Press, 2002.

\bibitem{HB.PLC.book}
H.~H. Bauschke and P.~L. Combettes, {\em Convex Analysis and Monotone Operator
  Theory in Hilbert Spaces}.
\newblock New York: Springer, 2011.

\bibitem{engel2004krls}
Y.~Engel, S.~Mannor, and R.~Meir, ``The kernel recursive least-squares
  algorithm,'' {\em IEEE Transactions on Signal Processing}, vol.~52, no.~8,
  pp.~2275--2285, 2004.

\bibitem{lagoudakis2003lspi}
M.~G. Lagoudakis and R.~Parr, ``Least-squares policy iteration,'' {\em J.\
  Mach.\ Learn.\ Res.}, vol.~4, pp.~1107--1149, Dec. 2003.

\bibitem{panaganti2022robust}
K.~Panaganti, Z.~Xu, D.~Kalathil, and M.~Ghavamzadeh, ``Robust reinforcement
  learning using offline data,'' {\em arXiv}, 2022.
\newblock abs/2208.05129.

\bibitem{Yuki:ICASSP23}
Y.~Akiyama, M.~Vu, and K.~Slavakis, ``Online and lightweight kernel-based
  approximate policy iteration for dynamic p-norm linear adaptive filtering.''
  Submitted for publication to arXiv, Oct. 2022.

\bibitem{RFF}
A.~Rahimi and B.~Recht, ``Random features for large-scale kernel machines,'' in
  {\em Proc.\ NIPS}, vol.~20, 2007.

\bibitem{ExperienceReplay}
T.~Schaul, J.~Quan, I.~Antonoglou, and D.~Silver, ``Prioritized experience
  replay,'' in {\em Proc.\ International Conference on Learning
  Representations}, 2016.

\bibitem{Bellemare:16}
M.~G. Bellemare, G.~Ostrovski, A.~Guez, P.~Thomas, and R.~Munos, ``Increasing
  the action gap: {N}ew operators for reinforcement learning,'' in {\em Proc.\
  AAAI Conference on Artificial Intelligence}, vol.~30, 2016.

\bibitem{Bartle.book:95}
R.~G. Bartle, {\em The Elements of Integration and Lebesgue Measure}.
\newblock John Wiley \& Sons, 1995.

\bibitem{bezanson2017julia}
J.~Bezanson, A.~Edelman, S.~Karpinski, and V.~B. Shah, ``Julia: A fresh
  approach to numerical computing,'' {\em SIAM Review}, vol.~59, no.~1,
  pp.~65--98, 2017.

\end{thebibliography}


\end{document}